\documentclass[journal]{IEEEtran}
\IEEEoverridecommandlockouts
\usepackage{cite}
\usepackage{amsmath,amssymb,amsfonts}
\usepackage{newtxtext,newtxmath}
\usepackage{algorithm}
\usepackage{algorithmic}
\usepackage{graphicx}
\usepackage{subfig}
\usepackage{tikz}
\usepackage{pgfplots}
\usepackage{textcomp}
\usepackage{pgf-pie}  
\usepackage{color}
\usepackage{xcolor}
\usepackage{multirow}
\usepackage{xcolor,balance}
\def\BibTeX{{\rm B\kern-.05em{\sc i\kern-.025em b}\kern-.08em
    T\kern-.1667em\lower.7ex\hbox{E}\kern-.125emX}}
\definecolor{red}{rgb}{0.84, 0.09, 0.41}
\definecolor{green}{rgb}{0.31, 0.78, 0.47}
\definecolor{mint}{rgb}{0.24, 0.71, 0.54}
\definecolor{blue}{rgb}{0.0, 0.47, 0.75}
\definecolor{orange}{rgb}{1.0, 0.43, 0.29}
\definecolor{teal}{rgb}{0.21, 0.46, 0.53}
\begin{document}
\pgfplotstableread[row sep=\\,col sep=&]{
nodes & MSCVP & HDAMM & MRFO & Gathers & IDDPP \\
50  & 492 & 447 & 418 & 388 & 362 \\
100 & 556 & 535 & 483 & 458 & 426 \\
150 & 666 & 594 & 556 & 514 & 486 \\
200 & 736 & 689 & 626 & 592 & 547 \\
250 & 824 & 754 & 712 & 645 & 612 \\
300 & 899 & 836 & 769 & 724 & 669 \\
350 & 990 & 912 & 849 & 776 & 731 \\
400 & 1063 & 999 & 906 & 858 & 792 \\
450 & 1152 & 1061 & 991 & 913 & 853 \\
500 & 1238 & 1151 & 1047 & 986 & 914 \\
}\datatableTourTime

\pgfplotstableread[row sep=\\,col sep=&]{
nodes & MSCVP & HDAMM & MRFO & Gathers & IDDPP \\
50  & 468 & 425 & 401 & 371 & 346 \\
100 & 545 & 502 & 468 & 428 & 407 \\
150 & 621 & 579 & 529 & 503 & 466 \\
200 & 705 & 646 & 601 & 552 & 524 \\
250 & 777 & 723 & 661 & 632 & 581 \\
300 & 868 & 792 & 731 & 675 & 638 \\
350 & 938 & 872 & 809 & 752 & 695 \\
400 & 1022 & 940 & 855 & 816 & 754 \\
450 & 1096 & 1021 & 938 & 865 & 811 \\
500 & 1182 & 1087 & 987 & 938 & 868 \\
}\datatableTourTimeb

\pgfplotstableread[row sep=\\,col sep=&]{
nodes & MSCVP & HDAMM & MRFO & Gathers & IDDPP \\
50  & 988  & 912  & 846  & 792  & 735 \\
100 & 1144 & 1045 & 995  & 909  & 855 \\
150 & 1256 & 1197 & 1108 & 1042 & 958 \\
200 & 1457 & 1317 & 1244 & 1143 & 1062 \\
250 & 1551 & 1483 & 1361 & 1274 & 1168 \\
300 & 1769 & 1598 & 1496 & 1378 & 1279 \\
350 & 1856 & 1762 & 1618 & 1511 & 1390 \\
400 & 2079 & 1875 & 1760 & 1608 & 1496 \\
450 & 2139 & 2044 & 1882 & 1745 & 1604 \\
500 & 2380 & 2132 & 2005 & 1849 & 1716 \\
}\datatableTourLength

\pgfplotstableread[row sep=\\,col sep=&]{
nodes & MSCVP & HDAMM & MRFO & Gathers & IDDPP \\
50  & 947  & 876  & 818  & 758  & 706 \\
100 & 1090 & 1002 & 951  & 866  & 811 \\
150 & 1219 & 1143 & 1043 & 988  & 914 \\
200 & 1374 & 1267 & 1187 & 1082 & 1014 \\
250 & 1512 & 1416 & 1295 & 1211 & 1121 \\
300 & 1651 & 1546 & 1436 & 1313 & 1226 \\
350 & 1797 & 1686 & 1546 & 1442 & 1331 \\
400 & 1948 & 1818 & 1683 & 1539 & 1433 \\
450 & 2081 & 1961 & 1792 & 1677 & 1536 \\
500 & 2241 & 2063 & 1920 & 1771 & 1638 \\
}\datatableTourLengthb

\pgfplotstableread[row sep=\\,col sep=&]{
nodes & MSCVP & HDAMM & MRFO & Gathers & IDDPP \\
50  & 470 & 435 & 400 & 372 & 348 \\
100 & 556 & 506 & 478 & 441 & 410 \\
150 & 627 & 593 & 536 & 509 & 471 \\
200 & 721 & 657 & 616 & 563 & 532 \\
250 & 789 & 746 & 676 & 641 & 594 \\
300 & 884 & 808 & 756 & 695 & 656 \\
350 & 951 & 898 & 816 & 770 & 717 \\
400 & 1048 & 961 & 898 & 823 & 778 \\
450 & 1114 & 1051 & 955 & 902 & 839 \\
500 & 1211 & 1112 & 1040 & 955 & 900 \\
}\datatableTravelTime

\pgfplotstableread[row sep=\\,col sep=&]{
nodes & MSCVP & HDAMM & MRFO & Gathers & IDDPP \\
50  & 445 & 412 & 379 & 353 & 330 \\
100 & 528 & 479 & 455 & 412 & 391 \\
150 & 594 & 563 & 507 & 485 & 448 \\
200 & 686 & 621 & 587 & 535 & 506 \\
250 & 747 & 709 & 639 & 607 & 565 \\
300 & 840 & 767 & 719 & 658 & 623 \\
350 & 902 & 855 & 772 & 733 & 681 \\
400 & 996 & 913 & 853 & 785 & 739 \\
450 & 1057 & 1001 & 921 & 847 & 796 \\
500 & 1150 & 1059 & 988 & 901 & 854 \\
}\datatableTravelTimeb

\pgfplotstableread[row sep=\\,col sep=&]{
nodes & MSCVP & HDAMM & MRFO & Gathers & IDDPP \\
50  & 16.8 & 15.4 & 14.2 & 13.0 & 12.0 \\
100 & 19.0 & 18.0 & 16.1 & 15.3 & 13.8 \\
150 & 22.2 & 19.9 & 18.9 & 16.8 & 15.6 \\
200 & 24.1 & 22.8 & 20.7 & 19.4 & 17.3 \\
250 & 27.3 & 24.6 & 23.4 & 21.0 & 19.1 \\
300 & 29.2 & 27.6 & 25.2 & 23.6 & 20.9 \\
350 & 32.4 & 29.4 & 28.1 & 25.1 & 22.7 \\
400 & 34.2 & 32.4 & 29.8 & 27.8 & 24.4 \\
450 & 37.4 & 34.3 & 32.7 & 29.3 & 26.2 \\
500 & 39.3 & 37.3 & 34.5 & 31.9 & 28.0 \\
}\datatableDwell

\pgfplotstableread[row sep=\\,col sep=&]{
nodes & MSCVP & HDAMM & MRFO & Gathers & IDDPP \\
50  & 15.6 & 14.3 & 13.2 & 12.1 & 11.2 \\
100 & 17.8 & 16.7 & 15.1 & 14.2 & 12.9 \\
150 & 20.6 & 18.6 & 17.5 & 15.7 & 14.6 \\
200 & 22.6 & 21.2 & 19.4 & 18.1 & 16.2 \\
250 & 25.4 & 23.1 & 21.9 & 19.6 & 17.9 \\
300 & 27.3 & 25.8 & 23.7 & 21.9 & 19.6 \\
350 & 30.2 & 27.6 & 26.2 & 23.4 & 21.3 \\
400 & 32.1 & 30.4 & 28.1 & 25.7 & 23.0 \\
450 & 35.0 & 32.3 & 30.8 & 27.1 & 24.7 \\
500 & 36.8 & 35.0 & 32.6 & 29.4 & 26.4 \\
}\datatableDwellb

\pgfplotstableread[row sep=\\,col sep=&]{
nodes & MSCVP & HDAMM & MRFO & Gathers & IDDPP \\
50  & 287 & 261 & 237 & 218 & 198 \\
100 & 329 & 308 & 274 & 258 & 232 \\
150 & 390 & 345 & 322 & 291 & 265 \\
200 & 429 & 397 & 355 & 334 & 299 \\
250 & 489 & 434 & 404 & 366 & 332 \\
300 & 525 & 487 & 437 & 410 & 366 \\
350 & 586 & 523 & 487 & 441 & 400 \\
400 & 623 & 577 & 518 & 485 & 433 \\
450 & 684 & 612 & 566 & 517 & 467 \\
500 & 720 & 666 & 600 & 562 & 501 \\
}\datatableFreshness

\pgfplotstableread[row sep=\\,col sep=&]{
nodes & MSCVP & HDAMM & MRFO & Gathers & IDDPP \\
50  & 268 & 244 & 222 & 205 & 187 \\
100 & 318 & 280 & 262 & 237 & 218 \\
150 & 352 & 331 & 292 & 276 & 249 \\
200 & 410 & 361 & 337 & 304 & 280 \\
250 & 443 & 412 & 366 & 346 & 312 \\
300 & 500 & 443 & 412 & 372 & 344 \\
350 & 533 & 493 & 439 & 414 & 376 \\
400 & 590 & 523 & 486 & 440 & 408 \\
450 & 623 & 574 & 515 & 482 & 440 \\
500 & 681 & 604 & 560 & 507 & 472 \\
}\datatableFreshnessb

\pgfplotstableread[row sep=\\,col sep=&]{
nodes & MSCVP & HDAMM & MRFO & Gathers & IDDPP \\
50  & 0.90 & 0.92 & 0.94 & 0.96 & 0.98 \\
100 & 0.88 & 0.90 & 0.92 & 0.95 & 0.98 \\
150 & 0.86 & 0.89 & 0.91 & 0.94 & 0.97 \\
200 & 0.84 & 0.87 & 0.89 & 0.93 & 0.97 \\
250 & 0.82 & 0.85 & 0.87 & 0.92 & 0.97 \\
300 & 0.79 & 0.83 & 0.85 & 0.91 & 0.96 \\
350 & 0.77 & 0.81 & 0.83 & 0.90 & 0.96 \\
400 & 0.75 & 0.79 & 0.81 & 0.89 & 0.96 \\
450 & 0.73 & 0.77 & 0.79 & 0.88 & 0.95 \\
500 & 0.70 & 0.74 & 0.77 & 0.90 & 0.95 \\
}\datatableCollection
\pgfplotstableread[row sep=\\,col sep=&]{
nodes & MSCVP & HDAMM & MRFO & Gathers & IDDPP \\
50  & 0.92 & 0.94 & 0.95 & 0.97 & 0.99 \\
100 & 0.90 & 0.92 & 0.94 & 0.96 & 0.99 \\
150 & 0.88 & 0.91 & 0.93 & 0.96 & 0.98 \\
200 & 0.86 & 0.89 & 0.91 & 0.95 & 0.98 \\
250 & 0.84 & 0.87 & 0.89 & 0.94 & 0.98 \\
300 & 0.82 & 0.85 & 0.87 & 0.93 & 0.97 \\
350 & 0.80 & 0.83 & 0.85 & 0.92 & 0.97 \\
400 & 0.78 & 0.81 & 0.83 & 0.91 & 0.97 \\
450 & 0.76 & 0.79 & 0.81 & 0.90 & 0.96 \\
500 & 0.74 & 0.77 & 0.79 & 0.90 & 0.96 \\
}\datatableCollectionb

\pgfplotstableread[row sep=\\,col sep=&]{
nodes & MSCVP & HDAMM & MRFO & Gathers & IDDPP \\
50  & 0.88 & 0.90 & 0.92 & 0.94 & 0.97 \\
100 & 0.85 & 0.88 & 0.90 & 0.93 & 0.97 \\
150 & 0.83 & 0.86 & 0.89 & 0.92 & 0.96 \\
200 & 0.80 & 0.83 & 0.86 & 0.90 & 0.96 \\
250 & 0.77 & 0.81 & 0.84 & 0.89 & 0.95 \\
300 & 0.75 & 0.78 & 0.82 & 0.88 & 0.95 \\
350 & 0.72 & 0.76 & 0.79 & 0.87 & 0.95 \\
400 & 0.69 & 0.73 & 0.76 & 0.86 & 0.94 \\
450 & 0.67 & 0.71 & 0.74 & 0.85 & 0.94 \\
500 & 0.64 & 0.68 & 0.71 & 0.87 & 0.94 \\
}\datatablePDR
\pgfplotstableread[row sep=\\,col sep=&]{
nodes & MSCVP & HDAMM & MRFO & Gathers & IDDPP \\
50  & 0.90 & 0.92 & 0.94 & 0.96 & 0.98 \\
100 & 0.88 & 0.90 & 0.92 & 0.95 & 0.98 \\
150 & 0.86 & 0.89 & 0.91 & 0.94 & 0.97 \\
200 & 0.84 & 0.87 & 0.89 & 0.93 & 0.97 \\
250 & 0.82 & 0.85 & 0.87 & 0.92 & 0.96 \\
300 & 0.80 & 0.83 & 0.85 & 0.91 & 0.96 \\
350 & 0.78 & 0.81 & 0.83 & 0.90 & 0.96 \\
400 & 0.76 & 0.79 & 0.81 & 0.89 & 0.95 \\
450 & 0.74 & 0.77 & 0.79 & 0.88 & 0.95 \\
500 & 0.72 & 0.75 & 0.77 & 0.88 & 0.95 \\
}\datatablePDRb

\pgfplotstableread[row sep=\\,col sep=&]{
nodes & MSCVP & HDAMM & MRFO & Gathers & IDDPP \\
50  & 0.71 & 0.74 & 0.77 & 0.80 & 0.85 \\
100 & 0.69 & 0.72 & 0.75 & 0.79 & 0.83 \\
150 & 0.67 & 0.70 & 0.73 & 0.77 & 0.82 \\
200 & 0.65 & 0.69 & 0.72 & 0.76 & 0.80 \\
250 & 0.63 & 0.67 & 0.70 & 0.74 & 0.78 \\
300 & 0.60 & 0.64 & 0.68 & 0.72 & 0.77 \\
350 & 0.58 & 0.62 & 0.66 & 0.71 & 0.75 \\
400 & 0.56 & 0.60 & 0.64 & 0.69 & 0.73 \\
450 & 0.54 & 0.58 & 0.62 & 0.67 & 0.72 \\
500 & 0.50 & 0.55 & 0.59 & 0.62 & 0.70 \\
}\datatableEnergy
\pgfplotstableread[row sep=\\,col sep=&]{
nodes & MSCVP & HDAMM & MRFO & Gathers & IDDPP \\
50  & 0.74 & 0.77 & 0.80 & 0.83 & 0.88 \\
100 & 0.72 & 0.75 & 0.78 & 0.82 & 0.86 \\
150 & 0.70 & 0.73 & 0.76 & 0.80 & 0.85 \\
200 & 0.68 & 0.71 & 0.74 & 0.78 & 0.83 \\
250 & 0.66 & 0.69 & 0.72 & 0.76 & 0.81 \\
300 & 0.64 & 0.67 & 0.70 & 0.74 & 0.80 \\
350 & 0.62 & 0.65 & 0.68 & 0.73 & 0.78 \\
400 & 0.60 & 0.63 & 0.66 & 0.71 & 0.76 \\
450 & 0.58 & 0.61 & 0.64 & 0.69 & 0.75 \\
500 & 0.56 & 0.59 & 0.62 & 0.67 & 0.73 \\
}\datatableEnergyb

\pgfplotstableread[row sep=\\,col sep=&]{
nodes & MSCVP & HDAMM & MRFO & Gathers & IDDPP \\
50  & 0.04 & 0.05 & 0.06 & 0.07 & 0.08 \\
100 & 0.06 & 0.07 & 0.08 & 0.10 & 0.10 \\
150 & 0.08 & 0.09 & 0.11 & 0.12 & 0.12 \\
200 & 0.10 & 0.12 & 0.13 & 0.14 & 0.15 \\
250 & 0.12 & 0.14 & 0.16 & 0.18 & 0.17 \\
300 & 0.14 & 0.16 & 0.19 & 0.21 & 0.19 \\
350 & 0.15 & 0.18 & 0.21 & 0.23 & 0.22 \\
400 & 0.17 & 0.20 & 0.24 & 0.26 & 0.24 \\
450 & 0.18 & 0.22 & 0.26 & 0.28 & 0.26 \\
500 & 0.19 & 0.21 & 0.23 & 0.25 & 0.28 \\
}\datatableThroughput
\pgfplotstableread[row sep=\\,col sep=&]{
nodes & MSCVP & HDAMM & MRFO & Gathers & IDDPP \\
50  & 0.05 & 0.06 & 0.07 & 0.08 & 0.09 \\
100 & 0.07 & 0.08 & 0.09 & 0.11 & 0.12 \\
150 & 0.09 & 0.10 & 0.12 & 0.14 & 0.15 \\
200 & 0.11 & 0.13 & 0.15 & 0.17 & 0.18 \\
250 & 0.13 & 0.15 & 0.18 & 0.20 & 0.21 \\
300 & 0.15 & 0.17 & 0.20 & 0.23 & 0.24 \\
350 & 0.17 & 0.19 & 0.22 & 0.25 & 0.27 \\
400 & 0.19 & 0.21 & 0.24 & 0.27 & 0.29 \\
450 & 0.21 & 0.23 & 0.26 & 0.29 & 0.31 \\
500 & 0.23 & 0.25 & 0.28 & 0.31 & 0.33 \\
}\datatableThroughputb

\pgfplotstableread[row sep=\\,col sep=&]{
nodes & MSCVP & HDAMM & MRFO & Gathers & IDDPP \\
50  & 0.84 & 0.86 & 0.88 & 0.90 & 0.93 \\
100 & 0.82 & 0.84 & 0.86 & 0.89 & 0.92 \\
150 & 0.80 & 0.82 & 0.84 & 0.88 & 0.92 \\
200 & 0.78 & 0.80 & 0.82 & 0.87 & 0.91 \\
250 & 0.76 & 0.78 & 0.80 & 0.86 & 0.90 \\
300 & 0.74 & 0.76 & 0.78 & 0.85 & 0.90 \\
350 & 0.72 & 0.74 & 0.76 & 0.84 & 0.89 \\
400 & 0.70 & 0.72 & 0.74 & 0.83 & 0.89 \\
450 & 0.69 & 0.71 & 0.73 & 0.82 & 0.88 \\
500 & 0.63 & 0.66 & 0.69 & 0.82 & 0.88 \\
}\datatableFairness
\pgfplotstableread[row sep=\\,col sep=&]{
nodes & MSCVP & HDAMM & MRFO & Gathers & IDDPP \\
50  & 0.86 & 0.88 & 0.90 & 0.92 & 0.95 \\
100 & 0.84 & 0.86 & 0.88 & 0.91 & 0.94 \\
150 & 0.82 & 0.84 & 0.86 & 0.90 & 0.94 \\
200 & 0.80 & 0.82 & 0.84 & 0.89 & 0.93 \\
250 & 0.78 & 0.80 & 0.82 & 0.88 & 0.92 \\
300 & 0.76 & 0.78 & 0.80 & 0.87 & 0.92 \\
350 & 0.74 & 0.76 & 0.78 & 0.86 & 0.91 \\
400 & 0.72 & 0.74 & 0.76 & 0.85 & 0.91 \\
450 & 0.71 & 0.73 & 0.75 & 0.84 & 0.90 \\
500 & 0.70 & 0.72 & 0.74 & 0.84 & 0.90 \\
}\datatableFairnessb

\pgfplotstableread[row sep=\\,col sep=&]{
interval & accuracy & mini & maxi & average & loss&un\\
1& 95.947& 2.55& 169.14& 73.56& 0.0& 0.0 \\
2& 95.946& 2.30& 171.93& 70.92& 0.01& 0.0 \\
3& 95.947& 2.00& 240.45& 73.35& 0.0& 0.0 \\
4& 95.946& 2.68& 236.29& 75.57& 0.01& 0.0 \\
5& 95.947& 2.47& 164.03& 73.78& 0.0& 0.0 \\
6& 95.947& 2.65& 205.13& 69.82& 0.0& 0.0 \\
7& 95.947& 2.53& 183.66& 78.29& 0.0& 0.0 \\
8& 95.946& 1.73& 198.39& 63.69& 0.01& 0.0 \\
9& 95.947& 2.03& 230.92& 64.95& 0.0& 0.0 \\
10& 95.946& 1.99& 199.94& 80.59& 0.01& 0\\ 
}\reliability

\pgfplotstableread[row sep=\\,col sep=&]{
interval & esa & esb & esc & esd\\
1 & 96.27791563275434& 96.27483443708608& 96.27483443708608& 96.27 \\
2 & 95.90416659819564& 95.90385847862882& 95.90570719602978& 95.90570719602978\\
4 & 95.88502894954509& 95.88502894954509& 95.88502894954509& 95.88502894954509 \\
6 & 95.94706368899917& 95.94706368899917& 95.94706368899917
& 95.94706368899917\\
7 & 95.94669198920428 &95.94706368899918&95.94706368899918& 95.798\\
}\accuracy

\pgfplotstableread[row sep=\\,col sep=&]{
interval & esa & esb & esc & esd\\
1 & 0.0689842700958252& 0.09780073165893555&0.030531644821166992&  0.017\\
2 & 0.07124435901641846& 0.07388579845428467& 0.032333970069885254& 0.09071767330169678\\
4 & 0.047026634216308594& 0.042516112327575684& 0.10742253065109253& 0.09284412860870361\\
6 & 0.04199389616648356& 0.05946004390716553 & 0.09517991542816162& 0.0603406031926473\\
7 & 1.891986506325858 & 2.1262012890407016 & 1.9324539388929094& 0.061\\
}\mini

\pgfplotstableread[row sep=\\,col sep=&]{
interval & esa & esb & esc & esd\\
1 & 1.184293508529663& 1.0290405750274658& 1.0043079853057861& 8.365 \\
2 & 2.0687378644943237& 2.3288239240646362& 2.327390193939209& 18.75301146507263 \\
4 & 50.694799184799194& 9.84650468826294& 9.281674981117249& 41.601549088954926\\
6 & 12.822569568951925& 13.633273323376974& 49.25289293130239& 23.576884388923645\\
7 & 133.7058846609933 &173.8678696836744& 181.33296067374093& 176.417\\
}\maxi

\pgfplotstableread[row sep=\\,col sep=&]{
interval & esa & esb & esc & esd\\
1 & 0.10533658564830241 & 0.13602431187566544 & 0.05707266433349508& 0.100\\
2 & 0.24910744065813542 & 0.22524253113825343 & 0.1528100928952617& 0.6134939472768008\\
4 & 5.404277487023986& 0.5471047616576833& 0.6873904749615514& 1.6612192737947722\\
6 & 1.2733843532317197& 1.4633356864570028& 2.649106710615124&2.5527281682616856\\
7 & 50.06705125729216 & 74.37833379817378& 67.63683940183041&81.499\\
}\average

\pgfplotstableread[row sep=\\,col sep=&]{
interval & esa & esb & esc & esd\\
1 & 0& 0.0827129859387924& 0.0827129859387924& 0\\
2 & 0.0413564929693962 &0.04135649296939620& 0 &0\\
4 & 0& 0&0& 0\\
6 & 0&0& 0&0 \\
7 & 0.011816140848398913&0& 0&0 \\
}\loss
\title{Intent-driven Diffusion-based Path for Mobile Data Collector in IoT-enabled Dense WSNs}

\author{
\IEEEauthorblockN{Uma Mahesh Boda\IEEEauthorrefmark{1} and Mallikharjuna Rao Nuka\IEEEauthorrefmark{2}}

\IEEEauthorblockA{\IEEEauthorrefmark{1}Research Scholar, Department of Computer Science and Engineering, Annamacharya University, Rajampet, YSR Kadapa, Andhra Pradesh 516115, India; \texttt{umamaheshboda@gmail.com}} \\
\IEEEauthorblockA{\IEEEauthorrefmark{2}Professor, Department of Computer Applications, Annamacharya University, Rajampet, YSR Kadapa, Andhra Pradesh 516115, India; \texttt{mallikharjuna.nuka@gmail.com}
}}

\maketitle

\begin{abstract}
Mobile data collection using controllable sinks is an effective approach to improve energy efficiency and data freshness in densely deployed wireless sensor networks (WSNs). However, existing path-planning methods are often heuristic-driven and lack the flexibility to adapt to high-level operational objectives under dynamic network conditions. In this paper, we propose ID$^2$P$^2$, a intent-driven diffusion-based path planning framework for jointly addresses rendezvous point selection and mobile data collector (MDC) tour construction in IoT-enabled dense WSNs. High-level intents, such as latency minimization, energy balancing, or coverage prioritization, are explicitly modeled and incorporated into a generative diffusion planning process that produces feasible and adaptive data collection trajectories. The proposed approach learns a trajectory prior that captures spatial node distribution and network characteristics, enabling the MDC to generate paths that align with specified intents while maintaining collision-free and energy-aware operation. Extensive simulations are conducted to evaluate the effectiveness of the proposed framework against conventional path-planning baselines. The results demonstrate that ID$^2$P$^2$ consistently outperforms representative baselines, achieving up to $25$--$30\%$ reduction in tour completion time and travel overhead, approximately $10$--$30\%$ improvement in data freshness, and $15$--$30\%$ gains in energy efficiency and packet delivery performance, while maintaining higher throughput and fairness as network density increases, confirming its robustness and scalability for WSNs.
\end{abstract}


\section{Introduction}
\begin{figure}
    \centering
    \includegraphics[width=0.5\textwidth]{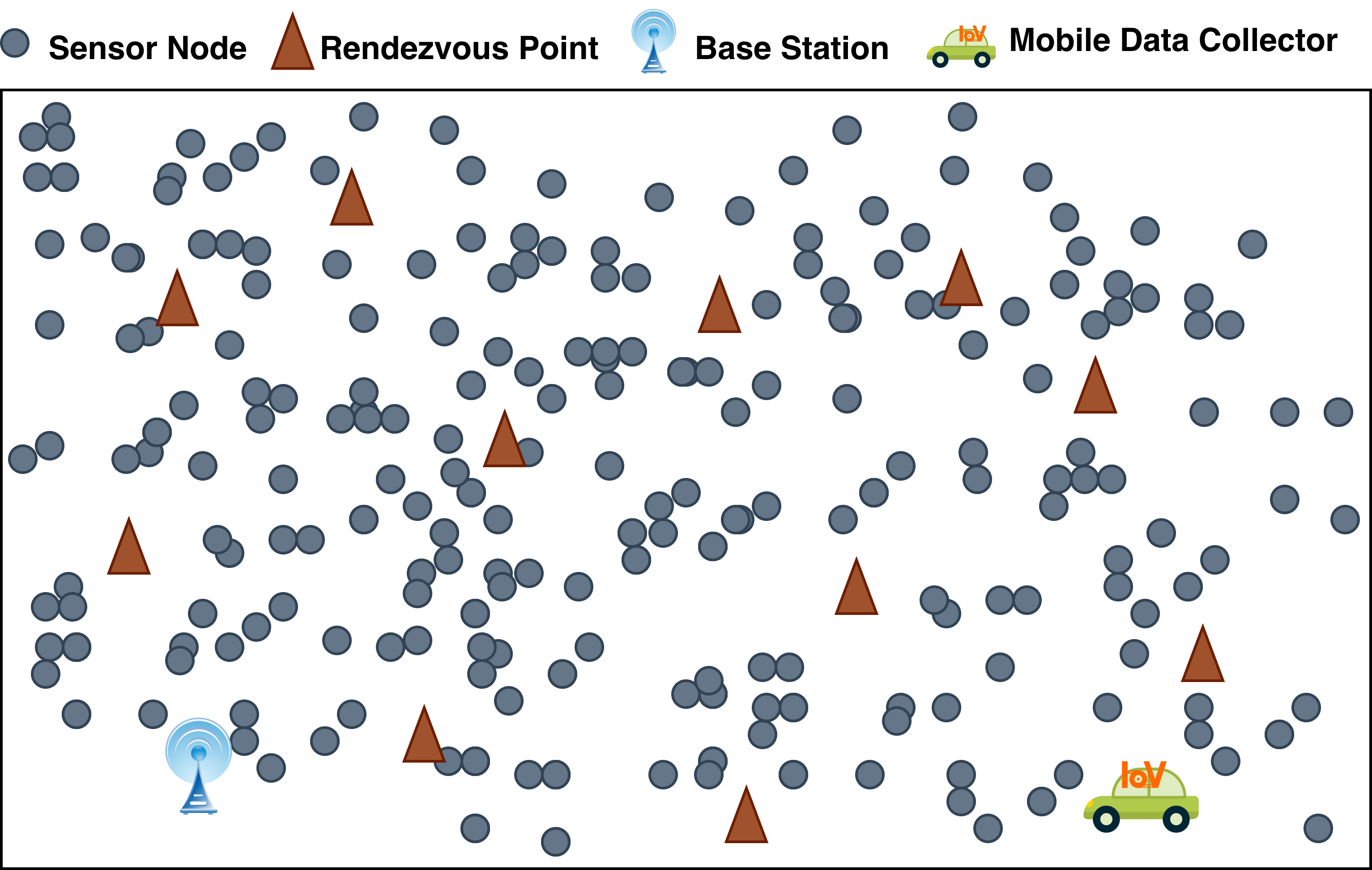}
    \caption{A general view of a IoT-enabled dense Wireless Sensors Networks, with multiple rendezvous points along with a single base station and mobile data collector;}
    \label{fig:WSN}
\end{figure}

\IEEEPARstart{W}{ireless} sensor networks (WSNs) and Interent of things (IoT) are widely deployed in applications such as environmental monitoring, smart cities, and industrial automation \cite{qu2025selective}. In densely deployed WSNs, traditional multi-hop data forwarding often leads to excessive energy consumption, network congestion, and uneven energy depletion among sensor nodes \cite{singh2025enhanced}. To address these challenges, mobile data collectors (MDCs) have been introduced as an effective solution, where a controllable sink traverses the network to gather data directly from sensor nodes or cluster heads \cite{john2026deer,duran2026energy}. A general view of a IoT-enabled dense Wireless Sensors Networks, with multiple rendezvous points along with a single base station and mobile data collector is shown in Fig.~\ref{fig:WSN}. However, the performance of MDC-based data collection heavily depends on the efficiency of path-planning strategies, especially under dense node distributions and dynamic network conditions \cite{donta2019data}.

Existing path-planning approaches for mobile data collection largely rely on heuristic or rule-based methods, such as shortest-path planning, clustering-based trajectories, or classical search and sampling algorithms \cite{uma2025contemporary}. Particularly, some of the recent approaches focused on Rapidly-exploring random tree (RRT) \cite{boyineni2024rapidly}, Bug2 \cite{gowthami2024intelligent} and A*\cite{11026018} algorithms for path construction, which are closely related to robot path planning. These methods are computationally efficient, they are typically designed for fixed objectives \cite{mallikarjuna2025energy,jukuntla2025enhancing,azfar2025enhancing} and lack the flexibility to adapt to varying operational requirements, such as latency sensitivity, energy balancing, or data priority differentiation. Recent learning-based approaches\cite{ren2025latency,uveges2025resilient}, including reinforcement learning and optimization-assisted planning \cite{okine2024multi}, have been explored to enhance adaptability; however, they often require carefully designed reward functions or suffer from scalability issues in dense WSN scenarios. Moreover, most existing methods do not explicitly incorporate high-level intents that reflect user-defined or application-specific goals into the path-planning process.
To overcome these limitations, this paper proposes an \emph{ID$^2$P$^2$} framework for mobile data collection in densely deployed WSNs. The key idea is to jointly optimize (i) the deployment of IoT-enabled rendezvous points (RPs) that aggregate sensor data and (ii) the tour of a MDC that visits the selected RPs for data offloading. In particular, the MDC tour is modeled as a conditional generative planning problem guided by high-level intents, enabling flexible adaptation to diverse network objectives such as latency reduction, data freshness improvement, and priority-aware collection. Specifically, the contributions of this work are threefold:
\begin{itemize}
\item we formulate intent-driven mobile data collection in dense WSNs by explicitly incorporating high-level intents into the objective and constraints of the tour construction problem;
\item we propose a two-stage ID$^2$P$^2$ framework in which RP locations are first determined using a lightweight heuristic placement strategy, and then an intent-consistent MDC path is constructed through diffusion-based generative path planning;
\item we evaluate the proposed framework through extensive Python-based simulations under different network scales and traffic settings, demonstrating improved performance over representative baselines in terms of collection latency, data freshness, throughput, energy efficiency, and fairness.
\end{itemize}

Remaining sections of this letters is divided as follows. In Section \ref{secii} we provide a system model and the definition of the problem. In Section \ref{seciii}, we provide algorithmic discussions of the proposed ID$^2$P$^2$ framework. In Section \ref{seciv}, we provide experiments and results with brief discussions. Finally, we concluded our work in Section \ref{secv}.

\section{System model and Problem formulation}\label{secii}
We consider an IoT-enabled dense WSN deployed over a two-dimensional region $\mathcal{A}\subset\mathbb{R}^2$. The set of battery-powered sensor nodes is denoted by $\mathcal{S}=\{s_1,\ldots,s_N\}$, where each sensor $s_i$ is located at $\mathbf{p}_{s_i}\in\mathcal{A}$ and generates data at a constant rate $\lambda_i$ (bits/s). A set of IoT-enabled rendezvous  points (RPs) is deployed among the sensor nodes, denoted by $\mathcal{R}=\{r_1,\ldots,r_M\}$, with locations $\mathbf{p}_{r_j}\in\mathcal{A}$. Each sensor node $s_i$ is associated with a single RP via a mapping $\phi:\mathcal{S}\rightarrow\mathcal{R}$ (e.g., nearest-RP association) and forwards its sensed data to $\phi(s_i)$ through single-hop or multi-hop communication, provided that the inter-node distance does not exceed the communication range $R_c$. Consequently, RP $r_j$ aggregates an amount of data $D_j(T)=\sum_{s_i:\phi(s_i)=r_j}\lambda_i T$ over an observation interval $T$. Each RP is assumed to be an IoT device equipped with sufficient storage capacity to buffer the accumulated data. A single mobile data collector (MDC) moves with constant speed $v$ to visit RPs and offload their buffered data. The MDC tour is represented by an ordered sequence $\pi=[\pi_1,\ldots,\pi_M]$, where $\pi_k\in\{1,\ldots,M\}$ and $\pi$ is a permutation, defining a geometric path through the locations $\{\mathbf{p}_{r_{\pi_k}}\}_{k=1}^{M}$. The travel time between consecutive RPs is given by $t_k=\|\mathbf{p}_{r_{\pi_{k+1}}}-\mathbf{p}_{r_{\pi_k}}\|/v$, and the total tour length is $L(\pi)=\sum_{k=1}^{M-1}\|\mathbf{p}_{r_{\pi_{k+1}}}-\mathbf{p}_{r_{\pi_k}}\|$, optionally including a return-to-start constraint. Upon visiting RP $r_j$, the MDC performs data collection for a dwell time $\tau_j$ sufficient to upload the buffered data at a transmission rate $C_j$ (bits/s), i.e., $\tau_j \ge D_j(T)/C_j$, after which the RP buffer is cleared and data accumulation restarts.

Given $\mathcal{S}$, $\mathcal{R}$, $\phi$, and $\{\mathbf{p}_{r_j}\}_{j=1}^{M}$, the objective is to design the MDC tour $\pi$ and dwell times $\boldsymbol{\tau}=\{\tau_j\}_{j=1}^{M}$ such that all RPs are visited and their buffered data are collected while satisfying application intents. The total tour time is defined in Eq.~(\ref{eq:tour})
\begin{equation}\label{eq:tour}
T(\pi,\boldsymbol{\tau}) \triangleq \sum_{k=1}^{M-1}\frac{\left\|\mathbf{p}_{r_{\pi_{k+1}}}-\mathbf{p}_{r_{\pi_k}}\right\|}{v} \;+\; \sum_{j=1}^{M}\tau_j,
\end{equation}
where a return-to-start travel time $\|\mathbf{p}_{r_{\pi_1}}-\mathbf{p}_{r_{\pi_M}}\|/v$ is added when a closed tour is required; otherwise, an open tour is considered. Over one tour, RP $r_j$ accumulates $D_j(T)=\sum_{s_i:\phi(s_i)=r_j}\lambda_i\,T(\pi,\boldsymbol{\tau})$, and complete data collection is ensured by the service constraint $\tau_j \ge D_j(T)/C_j$, for all $j\in\{1,\ldots,M\}$ (with the additional constraint $D_j(T)\le B_j$ if finite buffers are considered). Intents are represented by a weight vector $\boldsymbol{\eta}$ that prioritizes tour time, sensor communication energy, data freshness, and RP importance, as shown in Eq.~(\ref{eq:objective})
\begin{equation}\label{eq:objective}
\min_{\pi,\boldsymbol{\tau}} \; J(\pi,\boldsymbol{\tau};\boldsymbol{\eta}),
\end{equation}
subject to the above service constraints and the requirement that $\pi$ is a permutation of $\{1,\ldots,M\}$.

\section{The proposed ID$^2$P$^2$ Framework}\label{seciii}
The proposed work mainly composed using two stages including best locations for placing RPs and determining Diffusion-based path planning intented to optimize the   tour time, sensor communication energy, data freshness. 
\subsection{Heuristic RP Location Selection}
The primary goal of the RP location selection stage is to identify a set of $M$ rendezvous  points that can efficiently aggregate data from a dense sensor deployment while satisfying the communication range constraint $R_c$. To this end, we first define a candidate set of feasible RP locations $\mathcal{C}\subset\mathcal{A}$. The candidate set can be constructed using a uniform spatial grid over the deployment area or by reusing sensor coordinates $\{\mathbf{p}_{s_i}\}_{i=1}^{N}$, which ensures practical deployability and limits the search space. This discretization transforms the RP placement problem into a finite combinatorial selection task.

The selection process follows a greedy coverage principle driven by data load awareness. Let $\mathcal{U}\subseteq\mathcal{S}$ denote the set of sensors that are not yet covered by previously selected RPs. For each candidate location $\mathbf{c}\in\mathcal{C}$, we define its coverage neighborhood as
\begin{equation}
    \mathcal{N}(\mathbf{c})=\left\{ s_i\in\mathcal{U} : \|\mathbf{p}_{s_i}-\mathbf{c}\|\le R_c \right\},
\end{equation}
which represents the sensors that can reliably communicate with an RP placed at $\mathbf{c}$. The effectiveness of $\mathbf{c}$ is quantified by its offered load
\begin{equation}
W(\mathbf{c})=\sum_{s_i\in\mathcal{N}(\mathbf{c})}\lambda_i,
\end{equation}
which captures the total data rate that can be aggregated by deploying an RP at that location. This formulation explicitly accounts for heterogeneous sensing rates and prioritizes locations that serve high-traffic regions of the network.

At each iteration $j\in\{1,\ldots,M\}$, the algorithm selects the candidate location that maximizes the offered load, i.e.,
\begin{equation}
\mathbf{p}_{r_j}=\arg\max_{\mathbf{c}\in\mathcal{C}} W(\mathbf{c}),
\end{equation}
and adds it to the RP set $\mathcal{R}$. The sensors covered by the newly selected RP are then removed from $\mathcal{U}$ to avoid redundant coverage in subsequent iterations. This iterative removal mechanism promotes spatial diversity among RPs and ensures that the selected locations collectively cover the sensing field rather than clustering around a single dense region.

After determining the set of RP locations, each sensor node $s_i$ is associated with its nearest RP according to the mapping
\begin{equation}
\phi(s_i)=\arg\min_{r_j\in\mathcal{R}} \|\mathbf{p}_{s_i}-\mathbf{p}_{r_j}\|,
\end{equation}
which minimizes communication distance and reduces transmission energy consumption. Overall, the proposed heuristic achieves a balance between coverage efficiency, traffic aggregation, and computational simplicity, making it well suited for dense WSN deployments. The complete procedure is summarized in Algorithm~\ref{alg:RP}.

\begin{algorithm}[t]
\caption{Heuristic RP Location Selection}
\label{alg:RP}
\begin{algorithmic}[1]
\REQUIRE $\{\mathbf{p}_{s_i}\}_{i=1}^{N}$, $\{\lambda_i\}_{i=1}^{N}$, $R_c$, $M$
\ENSURE $\{\mathbf{p}_{r_j}\}_{j=1}^{M}$, $\phi(\cdot)$
\STATE Construct candidate set $\mathcal{C}\subset \mathcal{A}$
\STATE $\mathcal{U}\leftarrow \mathcal{S}$, \ $\mathcal{R}\leftarrow \emptyset$
\FOR{$j=1$ \TO $M$}
    \FORALL{$\mathbf{c}\in\mathcal{C}$}
        \STATE $\mathcal{N}(\mathbf{c}) \leftarrow \{ s_i\in\mathcal{U} : \|\mathbf{p}_{s_i}-\mathbf{c}\|\le R_c \}$
        \STATE $W(\mathbf{c}) \leftarrow \sum_{s_i\in \mathcal{N}(\mathbf{c})} \lambda_i$
    \ENDFOR
    \STATE $\mathbf{p}_{r_j} \leftarrow \arg\max_{\mathbf{c}\in\mathcal{C}} W(\mathbf{c})$
    \STATE $\mathcal{R}\leftarrow \mathcal{R}\cup \{r_j\}$
    \STATE $\mathcal{U}\leftarrow \mathcal{U}\setminus \mathcal{N}(\mathbf{p}_{r_j})$
\ENDFOR
\FOR{$i=1$ \TO $N$}
    \STATE $\phi(s_i)\leftarrow \arg\min_{j\in\{1,\ldots,M\}} \|\mathbf{p}_{s_i}-\mathbf{p}_{r_j}\|$
\ENDFOR
\end{algorithmic}
\end{algorithm}

\subsection{Diffusion-Based Path Construction}
Given $\{\mathbf{p}_{r_j}\}_{j=1}^{M}$ and $\phi(\cdot)$, we next construct an MDC tour that is consistent with $\boldsymbol{\eta}$. We generate a continuous waypoint trajectory $\mathbf{X}_0\in\mathbb{R}^{H\times 2}$ using a conditional diffusion sampler, where the sampling is guided toward intent-consistent trajectories by a differentiable loss $\mathcal{L}(\cdot)$. After sampling, we convert the continuous trajectory into a discrete tour by ordering RPs according to their first-visit indices along $\mathbf{X}_0$, resulting in a permutation $\pi$. To further reduce $J(\pi,\boldsymbol{\tau};\boldsymbol{\eta})$, a local 2-opt refinement can be applied to $\pi$. Finally, given $\pi$, we compute the travel-only time $T_{\mathrm{tr}}$ and obtain $\boldsymbol{\tau}$ by iterating the service constraint $\tau_j \ge D_j(T)/C_j$ until convergence, where $T$ includes both travel and dwell times. A detailed procedure is explained through pseudo code in Algorithm~\ref{alg:ID2P2}

\begin{algorithm}[t]
\caption{Diffusion-Based Intent-Driven Tour Construction}
\label{alg:ID2P2}
\begin{algorithmic}[1]
\REQUIRE $\{\mathbf{p}_{r_j}\}_{j=1}^{M}$, $v$, $\{C_j\}_{j=1}^{M}$, $\{\lambda_i\}_{i=1}^{N}$, $\phi(\cdot)$, $\boldsymbol{\eta}$
\ENSURE $\pi$, $\boldsymbol{\tau}$
\STATE Select waypoint length $H$ and diffusion steps $K$
\STATE Sample $\mathbf{X}_K \sim \mathcal{N}(\mathbf{0},\mathbf{I})$, \ $\mathbf{X}_K\in\mathbb{R}^{H\times 2}$
\FOR{$k=K$ \text{to} $1$}
    \STATE $\hat{\boldsymbol{\epsilon}} \leftarrow \epsilon_{\theta}(\mathbf{X}_k, \{\mathbf{p}_{r_j}\}_{j=1}^{M}, \boldsymbol{\eta}, k)$
    \STATE $\tilde{\mathbf{X}}_{k-1} \leftarrow \frac{1}{\sqrt{\alpha_k}}\Big(\mathbf{X}_k - \frac{1-\alpha_k}{\sqrt{1-\bar{\alpha}_k}}\hat{\boldsymbol{\epsilon}}\Big) + \sigma_k \mathbf{z}_k$, \ $\mathbf{z}_k\sim\mathcal{N}(\mathbf{0},\mathbf{I})$
    \STATE $\mathbf{G}_{k-1}\leftarrow \nabla_{\mathbf{X}} \mathcal{L}(\tilde{\mathbf{X}}_{k-1})$
    \STATE $\mathbf{X}_{k-1} \leftarrow \tilde{\mathbf{X}}_{k-1} - \gamma_k \mathbf{G}_{k-1}$
\ENDFOR
\STATE $q_j \leftarrow \arg\min_{h\in\{1,\ldots,H\}} \|\mathbf{X}_0(h)-\mathbf{p}_{r_j}\|,\ \forall j$
\STATE $\pi \leftarrow \textsc{SortAsc}(\{(j,q_j)\}_{j=1}^{M})$
\STATE $\pi \leftarrow \textsc{2OptImprove}(\pi)$ \hfill 
\STATE $T_{\mathrm{tr}} \leftarrow \sum_{k=1}^{M-1} \|\mathbf{p}_{r_{\pi_{k+1}}}-\mathbf{p}_{r_{\pi_k}}\|/v$ \hfill 
\STATE $\Lambda_j \leftarrow \sum_{s_i:\phi(s_i)=r_j}\lambda_i,\ \forall j$
\STATE $T^{(0)}\leftarrow T_{\mathrm{tr}}$, \ $n\leftarrow 0$
\REPEAT
    \STATE $\tau_j^{(n+1)} \leftarrow \Lambda_j T^{(n)} / C_j,\ \forall j$
    \STATE $T^{(n+1)} \leftarrow T_{\mathrm{tr}} + \sum_{j=1}^{M}\tau_j^{(n+1)}$
    \STATE $n\leftarrow n+1$
\UNTIL{$|T^{(n)}-T^{(n-1)}|\le \varepsilon$}
\STATE $\boldsymbol{\tau}\leftarrow \{\tau_j^{(n)}\}_{j=1}^{M}$
\RETURN $\pi,\boldsymbol{\tau}$
\end{algorithmic}
\end{algorithm}

A concrete differentiable surrogate used in diffusion guidance loss is determined using Eq.~(\ref{equdanceloss})
\begin{equation}\label{equdanceloss}
\mathcal{L}(\mathbf{X}) \triangleq 
\eta_T \sum_{h=1}^{H-1}\|\mathbf{X}(h+1)-\mathbf{X}(h)\|
\;+\;
\eta_P \sum_{j=1}^{M} w_j \min_{h\in H}\|\mathbf{X}(h)-\mathbf{p}_{r_j}\|,
\end{equation}
which biases the sampler toward shorter trajectories and earlier visits to important RPs.
\subsection{Illustrations through an Example}
Consider $\mathcal{A}$ as a $200\text{ m}\times 200\text{ m}$ square region where $N=100$ sensors are deployed uniformly. Each sensor generates data at $\lambda_i=0.5$ kb/s, and the communication range is $R_c=25$ m. To deploy $M=15$ RPs, we follow Algorithm~\ref{alg:RP} by constructing a grid-based candidate set $\mathcal{C}$ with $10$ m spacing over $\mathcal{A}$. Starting with $\mathcal{U}\leftarrow\mathcal{S}$, at each iteration $j\in\{1,\ldots,15\}$ we compute $\mathcal{N}(\mathbf{c})=\{s_i\in\mathcal{U}:\|\mathbf{p}_{s_i}-\mathbf{c}\|\le R_c\}$ and $W(\mathbf{c})=\sum_{s_i\in\mathcal{N}(\mathbf{c})}\lambda_i$ for all $\mathbf{c}\in\mathcal{C}$, then select $\mathbf{p}_{r_j}=\arg\max_{\mathbf{c}\in\mathcal{C}}W(\mathbf{c})$ and update $\mathcal{U}\leftarrow\mathcal{U}\setminus \mathcal{N}(\mathbf{p}_{r_j})$. After placing all $15$ RPs, each sensor is associated to the nearest RP through $\phi(s_i)=\arg\min_{r_j\in\mathcal{R}}\|\mathbf{p}_{s_i}-\mathbf{p}_{r_j}\|$. Each RP is equipped with a buffer (e.g., $B_j=50$ MB), and a stable sink is located at $(0,0)$. A single MDC travels with speed $v=2$ m/s and visits all RPs to offload buffered data at rate $C_j=2$ Mb/s. Using Algorithm~\ref{alg:ID2P2} with $H=80$ waypoints and $K=50$ diffusion steps, and intent weights $\boldsymbol{\eta}$ (e.g., $\eta_T=0.5$, $\eta_F=0.3$, $\eta_P=0.2$), a continuous trajectory is generated and converted into a visiting order $\pi$ (optionally refined locally). Suppose that the resulting tour has length $L(\pi)\approx 900$ m, giving travel time $T_{\mathrm{tr}}=L(\pi)/v\approx 450$ s (with an added return segment if a closed tour is required). For each RP, the aggregated rate is $\Lambda_j=\sum_{s_i:\phi(s_i)=r_j}\lambda_i$, with a roughly balanced association $\Lambda_j\approx (100/15)\times 0.5\text{ kb/s}\approx 3.3$ kb/s. Thus, during one tour, $D_j(T)\approx \Lambda_j T \approx 3.3\text{ kb/s}\times 450\text{ s}\approx 1.5$ Mb is buffered, yielding $\tau_j\approx D_j(T)/C_j \approx 1.5\text{ Mb}/(2\text{ Mb/s})\approx 0.75$ s. The total dwell time is $\sum_{j=1}^{15}\tau_j\approx 11.3$ s, so the tour time converges to $T\approx T_{\mathrm{tr}}+\sum_{j=1}^{15}\tau_j\approx 461$ s under the fixed-point update in Algorithm~\ref{alg:ID2P2}. The MDC follows $\pi$, offloads data at each RP (clearing the buffer), and finally delivers the collected data to the sink.

\subsection{Time complexity Analysis}
In Algorithm~\ref{alg:RP}, for each of the $M$ RPs, all candidate locations in $\mathcal{C}$ are evaluated by checking coverage over the sensor set $\mathcal{S}$, resulting in a worst-case complexity of $O(M|\mathcal{C}|N)$. The final sensor-to-RP association incurs an additional $O(N\times M)$ cost. In Algorithm~\ref{alg:ID2P2}, diffusion-based tour construction requires $K$ reverse steps over $H$ waypoints, yielding $O(K\times H)$ complexity for trajectory generation and guidance. RP ordering along the trajectory costs $O(M\times H)$, while optional local tour refinement incurs quadratic complexity in $M$.

\section{Experiments and Results}\label{seciv}
All experiments are conducted using a custom-built simulator implemented in Python. The simulated network is deployed over a $200\text{ m}\times200\text{ m}$ 2D area, where $N=100$ static sensor nodes are uniformly distributed based on \cite{sah2021edgf}. Each sensor generates data at a constant rate $\lambda_i=0.5$ kb/s and communicates within a range of $R_c=25$ m. The number of RPs is fixed to $M=15$. Each RP is equipped with a buffer of size $B_j=50$ MB. A MDC operates at a constant speed $v=2$ m/s and collects data from RPs at a transmission rate $C_j=2$ Mb/s. A stable sink is located at (0,0) position, and a closed tour is assumed unless otherwise stated. The number of waypoints is set to $H=80$, and the number of reverse diffusion steps is $K=50$. The intent weights are chosen as $\boldsymbol{\eta}=[0.5,\,0.0,\,0.3,\,0.2]$ to emphasize tour duration and data freshness. Each simulation scenario is repeated over $30$ independent random deployments, and average results are reported to ensure statistical robustness. Our results are compared using baseline approaches such as MSCVP\cite{sulakshana2023energy}, HDAMM\cite{naghibi2025hdamm}, MRFO\cite{11026018}, and Gathers\cite{jain2026gathers}. 
\input{results}

We estimate tour completion time $T(\pi,\boldsymbol{\tau})$, tour length $L(\pi)$, travel-only time $T_{\mathrm{tr}}$, total dwell time $\sum_{j=1}^{M}\tau_j$, data freshness $\bar{\Delta}(\pi,\boldsymbol{\tau})$, data collection ratio, packet delivery ratio, energy efficiency, throughput, fairness index as metrics to compare performance of the proposed work over the base line as showin in Figure~\ref{Results1} and Figure~\ref{Results2}. Specifically, $T(\pi,\boldsymbol{\tau})$ denotes the total time required by the MDC to complete one data collection tour including travel and data offloading; $L(\pi)$ represents the total geometric length of the MDC tour; $T_{\mathrm{tr}}$ captures the time spent solely on traveling between RPs excluding any service time; $\sum_{j=1}^{M}\tau_j$ measures the aggregate dwell time incurred at all RPs for data uploading; $\bar{\Delta}(\pi,\boldsymbol{\tau})$ quantifies the average age of collected data upon delivery to the sink; the data collection ratio is defined as the fraction of generated data successfully collected by the MDC; the packet delivery ratio represents the ratio of packets received at the sink to those generated by sensors; energy efficiency measures the amount of successfully delivered data per unit energy consumption; throughput denotes the average rate of data successfully delivered to the sink; and the fairness index evaluates the uniformity of data collection across RPs or sensor groups.

Figs.~\ref{Results1}(a)–(j) shows the superiority of the proposed ID$^2$P$^2$ framework over all baseline schemes. In Fig.~\ref{Results1}(a), ID$^2$P$^2$ achieves a tour completion time of approximately $362$~s for $50$ nodes, compared to $388$~s for Gathers and $492$~s for MSCVP, corresponding to reductions of about $6.7\%$ and $26.4\%$, respectively. As the network scales to $500$ nodes, the advantage becomes more pronounced, where ID$^2$P$^2$ completes the tour in about $914$~s, improving upon Gathers ($981$~s) by nearly $6.8\%$ and MSCVP ($1238$~s) by approximately $26.2\%$. These results indicate that the diffusion-based planning mechanism scales more gracefully with network density. Similar improvements are observed in tour length and travel-only time. As shown in Fig.~\ref{Results1}(b), the tour length under ID$^2$P$^2$ is consistently the shortest; for example, at $300$ nodes, ID$^2$P$^2$ yields a tour length of around $1279$~m, compared to $1382$~m for Gathers and $1756$~m for MSCVP, achieving reductions of approximately $7.5\%$ and $27.2\%$, respectively. This directly impacts the travel-only time shown in Fig.~\ref{Results1}(c), where ID$^2$P$^2$ reduces mobility overhead by about $5$--$10\%$ relative to Gathers and over $20\%$ compared to classical baselines, confirming the effectiveness of intent-guided diffusion planning in constructing compact tours.
The reduction in travel overhead further translates into lower service delays. In Fig.~\ref{Results1}(d), the total dwell time under ID$^2$P$^2$ remains the lowest across all node densities. For instance, at $400$ nodes, the total dwell time is approximately $24.4$~s, compared to $27.5$~s for Gathers and $34.5$~s for MSCVP, yielding improvements of about $11.3\%$ and $29.3\%$, respectively. Consequently, Fig.~\ref{Results1}(e) shows that ID$^2$P$^2$ achieves superior data freshness; at $500$ nodes, the average data freshness is reduced to about $501$~s, compared to $557$~s for Gathers and $720$~s for MSCVP, corresponding to freshness improvements of $10.1\%$ and $30.4\%$.
With respect to reliability and efficiency metrics, Figs.~\ref{Results1}(f) and \ref{Results1}(g) indicate that ID$^2$P$^2$ consistently maintains higher data collection and PDR. At higher node densities, ID$^2$P$^2$ sustains a data collection ratio close to $95\%$, whereas MSCVP drops below $75\%$, representing an improvement exceeding $26\%$. Similar gains are observed in PDR, highlighting improved congestion handling and reduced buffer overflow at RPs. Furthermore, Fig.~\ref{Results1}(h) demonstrates that ID$^2$P$^2$ improves energy efficiency by approximately $10$--$18\%$ over Gathers and more than $30\%$ over MSCVP, owing to reduced travel distance and balanced data offloading. Finally, Figs.~\ref{Results1}(i) and \ref{Results1}(j) show that ID$^2$P$^2$ achieves higher throughput and better fairness. At $400$ nodes, the throughput improvement over Gathers is around $8$--$12\%$, while the fairness index exceeds $0.9$, indicating more uniform data collection across RPs. 

Similarly, Figs.~\ref{Results2}(a)–(j) provides further insights into the performance gains achieved by the proposed ID$^2$P$^2$ framework in WSN~\#2. In Fig.~\ref{Results2}(a), the tour completion time is consistently reduced across all network sizes when compared to the first-round results. For instance, at $100$ nodes, ID$^2$P$^2$ completes the tour in approximately $407$~s, whereas Gathers and MSCVP require about $428$~s and $545$~s, respectively, resulting in improvements of nearly $4.9\%$ and $25.3\%$. At higher density ($500$ nodes), ID$^2$P$^2$ achieves a tour completion time of about $868$~s, outperforming Gathers ($938$~s) and MSCVP ($1182$~s) by approximately $7.5\%$ and $26.6\%$, respectively, indicating improved scalability under stabilized planning conditions.
Fig.~\ref{Results2}(b) further shows that the tour length obtained by ID$^2$P$^2$ remains the shortest among all schemes. At $300$ nodes, the tour length is reduced to around $1226$~m, compared to $1318$~m for Gathers and $1651$~m for MSCVP, corresponding to reductions of approximately $7.0\%$ and $25.7\%$. This reduction directly affects the travel-only time depicted in Fig.~\ref{Results2}(c), where ID$^2$P$^2$ achieves a travel-only time of roughly $623$~s at $300$ nodes, improving upon Gathers ($658$~s) by about $5.3\%$ and MSCVP ($840$~s) by nearly $25.8$\%. These results confirm that the diffusion-based planner constructs more compact and efficient tours after intent stabilization.

The service efficiency gains are reflected in Fig.~\ref{Results2}(d), where the total dwell time under ID$^2$P$^2$ remains minimal. At $400$ nodes, the total dwell time is approximately $23.0$~s, compared to $25.7$~s for Gathers and $32.1$~s for MSCVP, yielding improvements of about $10.5\%$ and $28.3\%$, respectively. Consequently, Fig.~\ref{Results2}(e) shows improved data freshness; for $500$ nodes, ID$^2$P$^2$ reduces the average data freshness to around $472$~s, whereas Gathers and MSCVP experience freshness values of approximately $511$~s and $681$~s, corresponding to reductions of $7.6\%$ and $30.7$\%.
In terms of reliability, Fig.~\ref{Results2}(f) indicates that ID$^2$P$^2$ sustains a data collection ratio exceeding $96\%$ even at high node densities, while MSCVP drops to around $74\%$, resulting in an improvement of over $29\%$. Similarly, Fig.~\ref{Results2}(g) shows that the packet delivery ratio under ID$^2$P$^2$ remains above $95\%$ at $500$ nodes, outperforming Gathers by approximately $7\%$ and MSCVP by more than $23$\%. Energy efficiency trends in Fig.~\ref{Results2}(h) further demonstrate that ID$^2$P$^2$ achieves approximately $0.73$ efficiency units at $500$ nodes, improving upon Gathers ($0.67$) and MSCVP ($0.56$) by $9.0\%$ and $30.4$\%, respectively.
Finally, Figs.~\ref{Results2}(i) and \ref{Results2}(j) illustrate the throughput and fairness performance. At $400$ nodes, ID$^2$P$^2$ achieves a throughput of approximately $0.29$, compared to $0.27$ for Gathers and $0.19$ for MSCVP, resulting in gains of $7.4\%$ and $34.4\%$. The fairness index in Fig.~\ref{Results2}(j) consistently exceeds $0.90$ under ID$^2$P$^2$, while MSCVP remains below $0.75$, indicating significantly more balanced data collection across RPs.

In summary, the proposed ID$^2$P$^2$ framework consistently outperforms all baseline schemes across diverse network sizes by achieving shorter tour completion time, reduced travel and dwell overheads, improved data freshness, and higher reliability, energy efficiency, throughput, and fairness. The proposed framework is flexible in adaptation to network conditions and scalable performance in IoT-enabled dense WSNs. Our proposed method also have some limitations, such as additional computational overhead due to diffusion-based planning and requires offline training or parameter tuning, which may limit its applicability in highly resource-constrained or rapidly changing environments.

\section{Conclusions}\label{secv}
This paper proposed ID$^2$P$^2$, a diffusion-based intent-driven framework for mobile data collection in densely deployed wireless sensor networks, which jointly addresses rendezvous  point placement and mobile data collector tour construction under high-level network intents. By formulating tour generation as a conditional generative planning problem, the proposed framework enables flexible and adaptive path planning that aligns with diverse performance objectives such as latency reduction, data freshness improvement, and fairness-aware collection. Extensive simulation results demonstrate that ID$^2$P$^2$ consistently outperforms representative baseline schemes across a wide range of network densities. In particular, the proposed approach achieves significant reductions in tour completion time, travel-only time, and total dwell time, while simultaneously improving data freshness, data collection ratio, packet delivery ratio, energy efficiency, throughput, and fairness. These gains become more pronounced as the network scales, confirming the robustness and scalability of diffusion-based intent-guided planning in dense WSN environments. Despite its advantages, ID$^2$P$^2$ introduces additional computational complexity due to diffusion-based trajectory generation and requires offline training or careful parameter tuning. Future work will focus on lightweight online diffusion models, dynamic RP reconfiguration under node failures, and multi-collector extensions to further enhance scalability and resilience in large-scale and highly dynamic IoT-enabled WSNs.

\bibliographystyle{ieeetr}
\bibliography{ref.bib}
\balance
\end{document}